\RequirePackage{fix-cm}
\documentclass[smallextended]{svjour3}       
\smartqed  
\usepackage{graphicx}
\usepackage{amsmath,revsymb,amssymb,adjustbox}
\begin{document}

\title{Coincidence Structures and Hard-Core Few-Body Interactions
}


\author{N.L.\ Harshman         \and
        Adam Knapp 
}


\institute{N.L.\ Harshman \at
  Department of Physics, American University, 4400 Massachusetts Ave.\ NW, Washington, DC 20016-8079, USA \\
              \email{harshman@american.edu}           
           \and
           Adam Knapp \at
              Department of Mathematics and Statistics, American University, 4400 Massachusetts Ave.\ NW, Washington, DC 20016-8050, USA \\
              \email{knapp@american.edu}  
}

\date{Received: date / Accepted: date}

\maketitle

\begin{abstract}
The symmetry and topology of the coincidence structure, i.e. the locus of points in configuration space corresponding to particles in the same position, plays a critical role in extracting universal properties for few-body models with hard-core interactions. The coincidence structure is a scale-invariant union of manifolds possessing rich symmetry. When there are zero-range hard-core two-body interactions, the coincidence structure forms a nodal surface for finite-energy wave functions in configuration space. More generally, it acts like a defect that changes the topology of configuration space in a way that depends on the dimension of the underlying space, the total number of particles, and the number of particles in the hard-core interaction. We show that for the specific case of three-body hard-core interactions in one-dimension, the configuration space is no longer simply-connected, providing a topological explanation for several models that exhibit anyonic behavior.
\end{abstract}

\section{Introduction}
\label{sect:intro}

\begin{figure}\centering
  \includegraphics[width=0.33\textwidth]{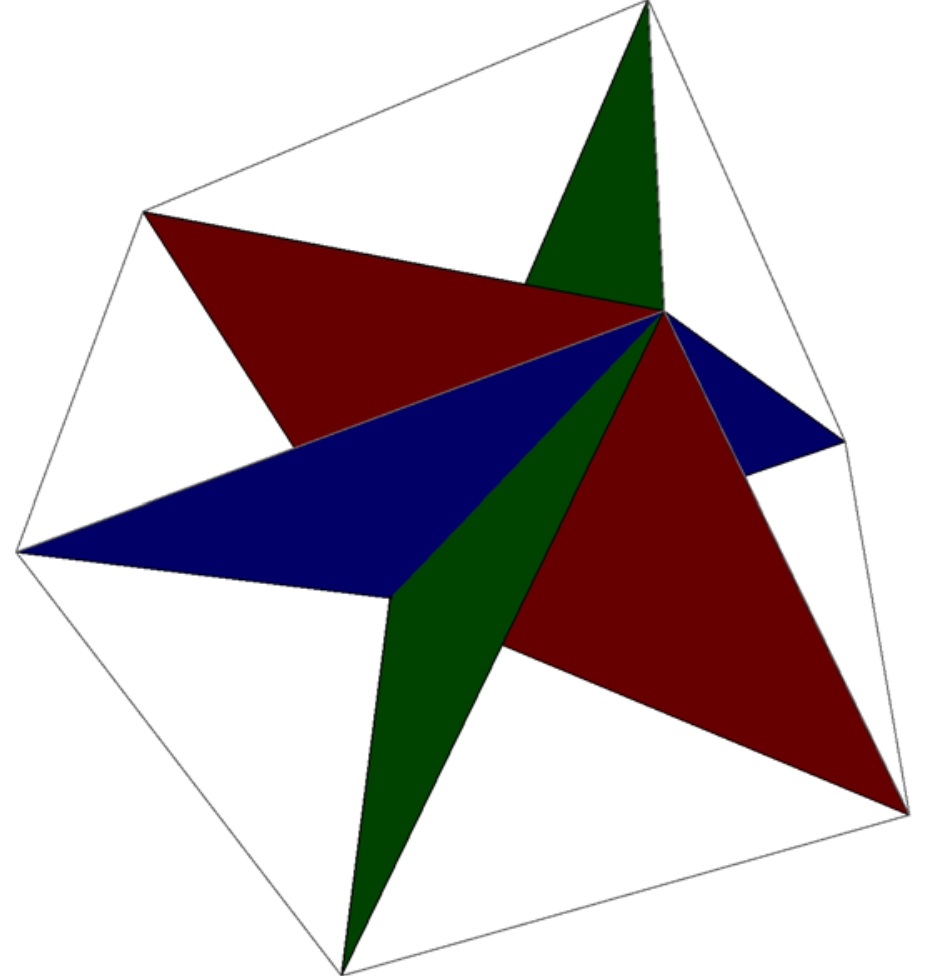}
\caption{This figure depicts the two-body coincidence structure $\mathcal{V}_2$ for three particles in one dimension (see text for notation). The three planes are $\mathcal{V}_{12}$ in red, $\mathcal{V}_{23}$ in green, and $\mathcal{V}_{13}$ in blue. The bounding box is arbitrary and breaks the translational symmetry along the line $\mathcal{V}_3$ where the three planes intersect at angles of $\pi/3$.}
\label{fig:AAA}       
\end{figure}

This article analyzes the properties of a geometrical object in configuration space called the coincidence structure. It is defined as the locus of points where two or more particles coincide. Because configuration space for $N$ particles in $d$ dimensions is $Nd$-dimensional, it is not an easy structure to depict except for low-dimensional cases. Fig.~\ref{fig:AAA} depicts an example with three particles in one dimension where the coincidence structure has the symmetry of a hexagonal prism and separates the configuration space into six disconnected sectors. 

One appeal of studying the coincidence structure is that it is a universal structure for few-body physics. Its symmetry and topology only depend on the number of the particles $N$ and the dimension of the space $d$. When the particles have different masses and complicated interactions, the coincidence structure does not necessarily have the same symmetry of the Hamiltonian, but the geometrical structure retains its same form. 

The coincidence structure is also particularly useful when the system has hard-core interactions. Hard-core interactions create forbidden regions or defects in configuration space, and two-body hard-core interactions define defects that have the same topology as the coincidence structure. In the limiting case of zero-range hard-core interactions, the coincidence structure also has the same symmetry in configuration space as the interaction. The coincidence structure is a nodal surface for all finite-energy solutions of the Schr\"odinger equation in configuration space.

The motivating application for this analysis of the coincidence structure is experiments with ultracold atoms in optical traps with tunable interactions~\cite{blume_few-body_2012}. Optical traps provide a flexible array of shapes that can approximate harmonic wells, double wells, infinite barriers, and lattices in one, two and three dimensions. In some experiments, the traps contain true few-body systems; they are populated by a deterministically-controlled number of atoms~\cite{serwane_deterministic_2011}. In others, the traps are filled with a gas of ultracold atoms and the underlying few-body interactions are probed by looking at dynamics of the gas, such as collective motion, trap lass, and coherence~\cite{bloch_many-body_2008}. Bright and dark solitons can also be treated as the effective interacting few-body systems~\cite{strecker_formation_2002,cornish_formation_2006}. The two-body interactions between atoms can be tuned to the hard-core limit via Feshbach resonances~\cite{chin_feshbach_2010}, possibly combined with confinement-induced resonances~\cite{olshanii_atomic_1998}. In addition to two-body hard-core interactions, we will also consider the case of hard-core three-body interactions (i.e.\ repulsive interactions when three particles coincide), and mechanisms to produce effective three-body interactions in ultracold atomic gases have been proposed~\cite{paredes_pfaffian-like_2007,mahmud_dynamically_2014,paul_hubbard_2016}.

In experiments with ultracold atoms, a driving impulse is the search for universality: few-body phenomena where characteristic scales drop out and the same dynamical structures are manifest from the nuclear to the molecular scale~\cite{braaten_universality_2006}. The coincidence structure is such a scale-free, universal feature.
Further, the topology of the coincidence structure, or more precisely, the topology of configuration space when the coincidence structure is removed, captures the essential, universal features of hard-core interactions. Two examples are well known: 
\begin{itemize}
\item In one dimension, two-body interactions partition configuration space into disconnected sectors where the particles are in a specific order. For hard-core, zero-range two-body interactions, the highly-symmetric nodal surface provided by the coincidence structure underlies Girardeau's famous solution for the fermionization of identical bosons~\cite{girardeau_relationship_1960}. This is an effect that is independent of the details of the trap shape and can be extended and generalized to multi-component fermions, bosons and their mixtures.
\item In two dimensions, two-body interactions do not segment configuration space into ordering sectors, but they do change the topology in a meaningful way. They create defects such that configuration space is no longer simply-connected, leading to the physics of anyons and fractional statistics~\cite{leinaas_theory_1977,wilczek_quantum_1982,viefers_ideal_1995}.
\end{itemize}

This article presents a third example where the topology of the coincidence structure becomes important: hard-core three-body interactions in one dimension. Similar to the case of two-body hard-core interactions in two-dimensions, the topology of configuration space is not simply-connected when the three-particle coincidence structure is removed. Fractional statistics and anyon-like behavior have been predicted for systems with hard-core, three-body interactions in one dimension~\cite{paredes_pfaffian-like_2007,girardeau_three-body_2010,keilmann_statistically_2011,paredes_non-abelian_2012,lange_strongly_2017}. By tracing these effects back to the topology of few-body configuration spaces, similarities with and differences between the cases of two-dimensional, two-body hard-core anyons and one-dimensional, three-body hard-core anyons come into clearer relief.

\section{Geometry of the Coincidence Structure}\label{sect:geo}

As a starting point for analysis, consider a model of $N$ particles moving and interacting in an underlying $d$-dimensional space $\mathcal{X}$. A typical situation is that underlying space is Euclidean $\mathcal{X}\sim\mathbb{R}^d$, although there are other physically interesting cases like rings $\mathbb{S}^1$ and spheres $\mathbb{S}^2$. Configuration space is the product of $N$ copies of the single-particle space $\mathcal{X}^{\times N}$. For Euclidean $\mathcal{X}$, the configuration space $\mathcal{X}^{\times N} =\mathbb{R}^{Nd}$ is again Euclidean.

Note that an alternate approach to handling indistinguishable particles is to define a reduced configuration space in which configurations that are only different by the exchange of indistinguishable particles are identified~\cite{leinaas_theory_1977}. Then the `true' configuration space is $\mathcal{X}^{\times N}/S_N$, where $S_N$ is the symmetric group of $N$ identical particles. Taking the quotient by the symmetric group changes the topology of the configuration space so it is no longer Euclidean. However, for now we will consider the `unquotiented' form of configuration space $\mathcal{X}^{\times N}$. One advantage is that this formulation also applies to non-identical particles, and even for identical particles it is often more convenient to solve the problem in $\mathcal{X}^{\times N}$ and then restrict to $\mathcal{X}^{\times N}/S_N$.

The $N$-body two-body coincidence structure $\mathcal{V}_2$ is the locus of all points in $\mathcal{X}^{\times N}$ where at least two particles are in the same position on the underlying space $\mathcal{X}$. For zero-range, two-body hard-core interactions, all these configurations are impossible and the coincidence structure is a nodal surface, i.e.\ all wave functions defined on $\mathcal{X}^{\times N}$ must vanish on $\mathcal{V}_2$.  

The two-body coincidence structure $\mathcal{V}_2$ is built as the union of all pairwise coincidence manifolds $\mathcal{V}_{ij}$ defined by ${\bf x}_i = {\bf x}_j$:
\begin{equation}
\mathcal{V}_2 = \bigcup_{\langle i,j \rangle}^N \mathcal{V}_{ij},
\end{equation}
where the union is over all pairs $\langle i,j \rangle$, $i \neq j$.
The constraint ${\bf x}_i = {\bf x}_j$ means the pair manifold $\mathcal{V}_{ij}$ has a dimension $Nd - d$, or more conveniently, it has a co-dimension $d$ that is same as the dimension of the underlying space $\mathcal{X}$. For a one-dimensional system $d=1$, that means there is one dimension perpendicular to each $\mathcal{V}_{ij}$ and so $\mathcal{V}_{ij}$ is a hypersurface that divides space into regions where either $x_i <x_j$ or $x_j < x_i$. For a two-dimensional system $d=2$, there are two dimensions perpendicular to $\mathcal{V}_{ij}$, analogous to a line in three-dimensional space. When the co-dimension is $d=3$ then $\mathcal{V}_{ij}$ is analogous a point in three-dimensional space.

Contained in the two-body coincidence structure $\mathcal{V}_2$ is the three-body coincidence structure $\mathcal{V}_3$, where three particles are in the same position on $\mathcal{X}$. The structure $\mathcal{V}_3$ can be built as the union of manifolds like $\mathcal{V}_{ijk}$ defined by ${\bf x}_i = {\bf x}_j = {\bf x}_k$ or alternatively as the intersection of two pairwise manifolds that share a particle:
\begin{equation}
\mathcal{V}_{ijk} = \mathcal{V}_{ij} \cap \mathcal{V}_{ik} = \mathcal{V}_{ij} \cap \mathcal{V}_{jk} = \mathcal{V}_{ik} \cap \mathcal{V}_{jk}.
\end{equation}
The manifold $\mathcal{V}_{ijk}$ has dimension $Nd - 2d$ and co-dimension $2d$.

Contained within $\mathcal{V}_3$ is the four-body coincidence structure $\mathcal{V}_4$, and so on
\begin{equation}
\mathcal{V}_2 \supset \mathcal{V}_3 \supset \mathcal{V}_4 \supset \cdots \supset \mathcal{V}_N
\end{equation}
until we get to the final $N$-body coincidence structure when all $N$ particles are at the same place in $\mathcal{X}$. Each $k$-body coincidence structure is built from the union of manifolds with co-dimension $(k-1)d$. If there are hard-core two-body interactions, then all elements $\mathcal{V}_k$ of this lattice of structures are nodal surfaces for configuration space wave functions. If there are hard-core three-body interactions and the two-body interactions are finite in strength, then configurations space wave functions must vanish only on  $\mathcal{V}_3$ through $\mathcal{V}_N$.

The $N$-body coincidence structure $\mathcal{V}_N$ is a manifold isomorphic to the underlying space $\mathcal{X}$. If the particles all have the same mass, then the manifold $\mathcal{V}_N$ is the center-of-mass degrees of freedom in configuration space. However, if the particles do not all have the same masses, then $\mathcal{V}_N$ and the center-of-mass structure are not aligned. The $(N-1)d$-dimensional space $\mathcal{V}^\perp_N$ that is orthogonal to $\mathcal{V}_N$ is a kind of relative configuration space that does not take into account different particle masses. For equal-mass particles it is the standard relative configuration space.
 
\section{Symmetries of the Coincidence Structure}

In this section, we consider what transformations of configuration space leave $\mathcal{V}_2$, or its substructures $\mathcal{V}_k$ for $2 < k \leq N$, invariant. For the one-dimensional case, the symmetries of $\mathcal{V}_2$ have been previously described in \cite{harshman_one-dimensional_2016,harshman_one-dimensional_2016-1}. The symmetries of $\mathcal{V}_k$ are the same as the symmetries of a few-body system with identical particles, zero-range $k$-body hard-core interactions, and no external trapping potential.

The most obvious set of symmetries is that $\mathcal{V}_2$ is invariant under the representation of the group of particle permutations $\mathrm{S}_N$ on $\mathcal{X}^{\times N}$. The group of transformations can be generated by the $N-1$ pairwise exchanges ${\bf x}_1 \leftrightarrow {\bf x}_2$ through ${\bf x}_{N-1} \leftrightarrow {\bf x}_N$. In the Euclidean case, each of these exchanges is realized by an orthogonal linear transformation $\sigma(ij)$ on $\mathbb{R}^{Nd}$ that inverts configuration space around $\mathcal{V}_{ij}$. When $d=1$, an `inversion' is a actually a reflection across the co-dimension $d=1$ hypersurface $\mathcal{V}_{ij}$. When $d=2$ and $d=3$, the inversion is a rotation around the co-dimension $d=2$ `hyperline' $\mathcal{V}_{ij}$ or a true inversion through the co-dimension $d=3$ `hyperpoint' $\mathcal{V}_{ij}$, respectively.

Under the exchange $\sigma(ij)$, the manifold $\mathcal{V}_{ij}$ is invariant, but the other manifolds corresponding to different pairs are permuted. Similarly the manifolds $\mathcal{V}_{ijk}$ are permuted by all permutations except the specific three-cycle $\sigma(ijk)$, but as with $\mathcal{V}_2$ the three-body coincidence structure $\mathcal{V}_3$ as a whole is invariant. So are the higher-body coincidence structures $\mathcal{V}_k$ for $k<N$. The $N$-body coincidence structure $\mathcal{V}_N$ is completely invariant under particle permutations.

Note that $S_N$ is a symmetry of all $\mathcal{V}_k$ even if the particles are not identical. If the particles all have the same mass, feel the same trap, and have the same interaction properties, then $S_N$ will also be a symmetry of the Hamiltonian describing the system. 

In addition to invariance under a representation of $\mathrm{S}_N$, all $\mathcal{V}_k$ are invariant under uniform translations ${\bf x}_i \rightarrow {\bf x}_i + {\bf a}$ for ${\bf a} \in \mathbb{R}^d$, uniform rotations and reflections  ${\bf x}_i \rightarrow O{\bf x}_i$ for $O\in \mathrm{O}(d)$, and their combinations. In other words, the coincidence structure has Euclidean symmetry $\mathrm{E}_d \sim \mathrm{O}(d) \ltimes \mathbb{R}^d$, the semidirect product of the orthogonal group and the translation group. Unlike particle permutations $\mathrm{S}_N$ which permute manifolds of $\mathcal{V}_k$, transformations in $\mathrm{E}_d$ map manifolds $\mathcal{V}_{ij}$, $\mathcal{V}_{ijk}$, etc.\ onto themselves.

Like particle permutation symmetry, the Euclidean symmetry $\mathrm{E}_d$ of the coincidence structures $\mathcal{V}_k$ is not necessarily a symmetry of the few-body Hamiltonian. For identical particles and no external potentials, then the Hamiltonian has this $\mathrm{E}_d$ symmetry.

When $N>2$, all $\mathcal{V}_k$ have an additional symmetry denoted: inversion in the $(N-1)d$-dimensional space $\mathcal{V}^\perp_N$ orthogonal to $\mathcal{V}_N$. Denote this symmetry transformation $i^\perp$ and the order-two group it generates $\mathrm{Z}^\perp \sim \mathrm{Z}_2$. For identical particles, $\mathcal{V}_N$ corresponds the center-of-mass degrees of freedom, the space $\mathcal{V}^\perp_N$ is the relative configuration space, and this inversion corresponds to a transformation that defines relative parity. For $N>2$, this symmetry cannot be generated from the transformations derived from the $\mathrm{S}_N$ symmetry of particular permutations or the $\mathrm{O}(d)$ symmetry of $\mathrm{E}_d$. Except for quadratic potentials and identical particles, in which case the Hamiltonian separates along the partition $\mathcal{V}_N \times \mathcal{V}^\perp_N$, relative parity is not a symmetry of the few-body Hamiltonian.

Combining particle permutation symmetry and Euclidean symmetry, each coincidence structure has at least the symmetry $\mathrm{S}_N \ltimes \mathrm{E}_d$, with an additional factor of $\mathrm{Z}^\perp$ when $N>2$. As an example, consider the coincidence structure when $d=1$ and the underlying space is $\mathcal{X} = \mathbb{R}$ For $N=2$, the structure $\mathcal{V}_2$ is a line in configuration space $\mathcal{X}^{\times 2} = \mathbb{R}^2$. The symmetries of a line include translations along the line $\mathcal{V}_2$, reflections across any line perpendicular to $\mathcal{V}_2$, and reflections across $\mathcal{V}_2$ itself. For $N=3$, the structure $\mathcal{V}_2$ is the intersection of three planes in configuration space $\mathcal{X}^{\times 3} = \mathbb{R}^3$  (see Fig.~\ref{fig:AAA}). The symmetry includes
\begin{itemize}
\item the Euclidean transformations in $\mathrm{E}_1$ realized by translations along the line $\mathcal{V}_3$ where the three planes intersect and reflections across the plane perpendicular to $\mathcal{V}_3$,
\item the particle permutations in $\mathrm{S}_3$ realized by the reflections across each of the planes and rotations by $\pi/3$ around $\mathcal{V}_3$,
\item the additional inversion $i^\perp$ realized by a rotation by $\pi$ around $\mathcal{V}_3$, and
\item all their combinations.
\end{itemize}
Combined, these are the symmetries of an infinitely-long hexagonal prism. Restricted to the space $\mathcal{V}^\perp_3$, these transformations form the point symmetry group $\mathrm{D}_6 \sim \mathrm{S}_3 \times \mathrm{Z}_2$ of a hexagon. Similarly, restricted to the space $\mathcal{V}^\perp_4$, the four-particle, two-body coincidence structure $\mathcal{V}_2$ for one-dimensional particles has the symmetry of a cube $\mathrm{O}_h \sim \mathrm{S}_4 \times \mathrm{Z}_2$ (see Fig.~\ref{fig:BBB}).

Note that $\mathcal{V}_3$ has additional symmetries beyond $\mathcal{V}_2$. For example, in the $N=3$, $d=1$ case depicted in Fig.~\ref{fig:AAA}, the structure $\mathcal{V}_3$ has the symmetry of a line in three-dimensions. A full accounting of the symmetries of the higher-order structures for arbitrary particle number and dimension, and extended to non-trivial base spaces $\mathcal{X}$, awaits attention.

\section{Separability of the Coincidence Structure}

There is an additional symmetry of the coincidence structure: it is scale invariant. Under a uniform scale transformation ${\bf x}_i \to s {\bf x}_i$ for any real $s$, the coincidence structure $\mathcal{V}_2$ and any substructures do not change their structure. For example, all angles are preserved. Further, that property also holds true for the restrictions of any $\mathcal{V}_k$ to the space $\mathcal{V}^\perp_N$. 

The symmetry analysis from the previous section already suggested that the separation of variables $\mathcal{X}^{\times N} = \mathcal{V}_N \times \mathcal{V}^\perp_N$ is useful. Scale invariance suggests further partitioning the relative configuration space $\mathcal{V}^\perp_N = \mathcal{R} \times \mathcal{A} \sim  \mathbb{R}_+ \times \mathbb{S}^{(N-1)d-1}$ when the underlying space is Euclidean.  This decomposes the relative degrees of freedom of the space $\mathcal{V}^\perp_N$ into a sphere $\mathcal{A} \sim \mathbb{S}^{(N-1)d-1}$ with $(N-1)d-1$ angles and  a radius $\rho$ in $\mathcal{R} \sim \mathbb{R}_+$. The radius $\rho$ can be expressed as
\begin{equation}
\rho^2  = \frac{1}{N} \left\{ (N-1)\sum_{i=1}^N {\bf x}_i\cdot{\bf x}_i - 2 \sum_{\langle i,j \rangle} {\bf x}_i \cdot {\bf x}_j  \right\}
\end{equation}
For identical particles the radius $\rho$ is the relative hyperradius and $\mathbb{S}^{(N-1)d-1}$ are the hyperangular degrees of freedom.

Combining Euclidean invariance and scale invariance means the coincidence structures $\mathcal{V}_k$ are independent of the degrees of freedom in $\mathcal{V}_N$ and  $\mathcal{R}$. Because the coincidence structure has no dependence on these degrees of freedom, the problem of solving for energy eigenstates of the Hamiltonian is reduced to solving the Schr\"odinger equation within sectors of the sphere $\mathbb{S}^{N(d-1) -1}$ with Dirichlet boundary conditions on the intersection of $\mathcal{V}_k$ with the sphere $\mathcal{A}$. This separability can be exploited to analyze Hamiltonians of identical particles with hard-core interactions in harmonic traps or free space, or for Hamiltonians of particles with different masses in free space or in equal-frequency traps. See \cite{harshman_integrable_2017} for a recent application of this separability to identify integrable mixed-mass systems with hard-core two-body interactions in one dimension.

\section{Topology of Configuration Space with the Coincidence Structure Removed}

The configuration space $\mathcal{X}^{\times N} = \mathbb{R}^{Nd}$ for $N$ particles in a $d$-dimensional Euclidean space  is topologically trivial:
\begin{itemize}
\item It is \emph{connected} (aka $0$-connected): all points can be connected by a path in the space.
\item It is \emph{simply connected} (or $1$-connected): all loops in the space can be contracted into a point.
\item Generally, it is $k$-connected: all $k$-spheres can be contracted to a point.
\end{itemize}
However, when there are $k$-body hard-core interactions, the topology becomes more complicated. 
Hard-core $k$-body interactions introduce defects with co-dimension $\bar{d}=(k-1)d$ and the topology of the remaining configuration space $\mathcal{X}_{N,d,k} \equiv \mathcal{X}^{\times N} - \mathcal{V}_k$ can become less connected.

\begin{figure*}
\centering
 \adjustbox{trim={.05\width} {.21\height} {0.1\width} {.4\height},clip}{\includegraphics[width= .95\textwidth]{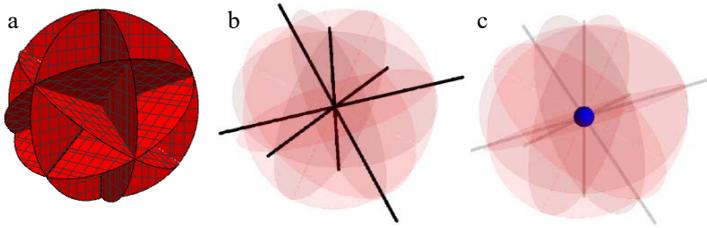}}
\caption{These three figures depict the relative configuration space $\mathcal{V}_4^\perp$ for four particles in one dimension. Each subfigure highlights the (a) two-body coincidence structure $\mathcal{V}_2$ in red planes; (b) three-body coincidence structure $\mathcal{V}_3$ in black lines; and (c) four-body coincidence structure $\mathcal{V}_4$ as a blue sphere. When there are hard-core two-body interactions, the structure $\mathcal{V}_2$ separates configuration space into 24 disconnected sectors. When there are hard-core three-body interactions, configuration space remains connected, but it is no longer simply-connected. There are loops around the structure $\mathcal{V}_3$ that cannot be contracted. }
\label{fig:BBB}       
\end{figure*}

Defects with $\bar{d} = 1$ make the configuration space disconnected. Co-dimension $\bar{d} = 1$ is only possible for the configuration space $\mathcal{X}_{N,1,2}$ of hard-core two-body interactions in one spatial dimension $d=1$. The $N(N-1)/2$ hyperplanes $\mathcal{V}_{ij}$ slice configuration space into $N!$ disconnected pieces. See Fig.~\ref{fig:BBB} for a depiction of $\mathcal{X}_{4,1,2}$. Each of these sectors  is dynamically isolated from the others and corresponds to a fixed order of the particles. Each of these $N!$ ordering sectors is identical and the permutation group of ordering sectors $\mathrm{S}_{N!}$ is a symmetry, and not just $\mathrm{S}_N$~\cite{harshman_identical_2017}. For equal mass particles, this is a symmetry of the Hamiltonian for any trap shape. Totally antisymmetrized wave functions must vanish on $\mathcal{V}_2$, and this underlies the famous Girardeau fermionization of hard-core contact-interaction bosons in one dimension~\cite{girardeau_relationship_1960}.

When $\bar{d} = 2$, configuration space remains connected, but not simply-connected. That means not every trajectory in configuration space can be continuously deformed into any other trajectory. Instead, there topological equivalence classes of trajectories whose structure depends on the geometry of the co-dimension $\bar{d}=2$ defect~\cite{viefers_ideal_1995}. The trajectories fall into equivalence classes that are described by the fundamental group $\pi_1$ (or first homotopy group) of the configuration space.

There are only two spaces $\mathcal{X}_{N,d,k}$ with $\bar{d} = 2$ defects from $k$-body hard-core interactions: $\mathcal{X}_{N,2,2}$ and $\mathcal{X}_{N,1,3}$. The first case was identified by Leinaas and Myrheim forty years ago~\cite{leinaas_theory_1977} and that paper is considered the starting point for fractional statistics and anyonic physics~\cite{wilczek_quantum_1982}. The fundamental group of $\mathcal{X}_{N,2,2}$ is the pure braid group $\mathrm{PB}_N$, and the fundamental group for configuration space quotiented by the symmetric group $\mathcal{X}_{N,2,2}/\mathrm{S}_N$ is the braid group $\mathrm{B}_N$~\cite{kassel_braid_2008}. The braid group has been studied exhaustively because of connections from everything from abstract knot theory to applications to quantum computing.

Apparently, the similar case of $\mathcal{X}_{N,1,3}$ is much less studied. In Fig.~\ref{fig:BBB}, the structure of $\mathcal{V}_3$ for four identical particles  restricted to relative configuration $\mathcal{V}^\perp_4$ is depicted. Contracting dimensions that are simply-connected, the topology of $\mathcal{X}_{4,1,3}$ is equivalent to the sphere $\mathcal{A} = \mathbb{S}^2$ punctured by eight holes. That structure is homotopy equivalent to the wedge product of seven circles $\bigvee_7 \mathbb{S}^1$ and the fundamental group is the free group with seven generators $\ast_7\mathbb{Z}$~\cite{harshman_notitle_nodate}.

By analogy with the braid groups, we propose the name \emph{pure traid group} $\mathrm{PT}_N$ for the fundamental group of $\mathcal{X}_{N,1,3}$ and \emph{traid group} $\mathrm{T}_N$ for the corresponding group of $\mathcal{X}_{N,1,3}/\mathrm{S}_N$. The traid groups do not seem to have been described before, but the rank of $\mathrm{PT}_N$ (equivalent to the Betti number $b_1$ of the configuration space $\mathcal{X}_{(N, 1, 3)}$) has been calculated in the context of graph theory and motivated by questions of computational complexity~\cite{bjorner_homology_1995}. Preliminary results suggest that the traid groups have representations that support abelain and non-abelian anyonic solutions~\cite{harshman_notitle_nodate}, and connecting these to previous results on anyons in one dimension with hard-core three-body interactions is a work in progress.

\begin{acknowledgements}
NLH would like the thank the Aarhus University Research Foundation for sabbatical support during the beginning of this project and D.\ Blume, G.\ Bruun, P.R.\ Johnson, N.J.S.\ Loft, J.M.\ Midtgaard and M.\ Olshanii for stimulating discussions and useful references.
\end{acknowledgements}



\begin{thebibliography}{10}
\providecommand{\url}[1]{{#1}}
\providecommand{\urlprefix}{URL }
\expandafter\ifx\csname urlstyle\endcsname\relax
  \providecommand{\doi}[1]{DOI \discretionary{}{}{}#1}\else
  \providecommand{\doi}{DOI \discretionary{}{}{}\begingroup
  \urlstyle{rm}\Url}\fi

\bibitem{blume_few-body_2012}
D.~Blume, Reports on Progress in Physics \textbf{75}(4), 046401 (2012).

\bibitem{serwane_deterministic_2011}
F.~Serwane, G.~Z{\"u}rn, T.~Lompe, T.B. Ottenstein, A.N. Wenz, S.~Jochim,
  Science \textbf{332}(6027), 336 (2011).

\bibitem{bloch_many-body_2008}
I.~Bloch, J.~Dalibard, W.~Zwerger, Rev. Mod. Phys. \textbf{80}(3), 885 (2008).

\bibitem{strecker_formation_2002}
K.E. Strecker, G.B. Partridge, A.G. Truscott, R.G. Hulet, Nature
  \textbf{417}(6885), 150 (2002).

\bibitem{cornish_formation_2006}
S.L. Cornish, S.T. Thompson, C.E. Wieman, Phys. Rev. Lett. \textbf{96}(17),
  170401 (2006).

\bibitem{chin_feshbach_2010}
C.~Chin, R.~Grimm, P.~Julienne, E.~Tiesinga, Rev. Mod. Phys. \textbf{82}(2),
  1225 (2010).

\bibitem{olshanii_atomic_1998}
M.~Olshanii, Phys. Rev. Lett. \textbf{81}(5), 938 (1998).

\bibitem{paredes_pfaffian-like_2007}
B.~Paredes, T.~Keilmann, J.I. Cirac, Phys. Rev. A \textbf{75}(5), 053611
  (2007).

\bibitem{mahmud_dynamically_2014}
K.W. Mahmud, E.~Tiesinga, P.R. Johnson, Physical Review A \textbf{90}(4)
  (2014).

\bibitem{paul_hubbard_2016}
S.~Paul, P.R. Johnson, E.~Tiesinga, Phys. Rev. A \textbf{93}(4), 043616 (2016).

\bibitem{braaten_universality_2006}
E.~Braaten, H.W. Hammer, Physics Reports \textbf{428}(5), 259 (2006).

\bibitem{girardeau_relationship_1960}
M.~Girardeau, Journal of Mathematical Physics \textbf{1}(6), 516 (1960).

\bibitem{leinaas_theory_1977}
J.M. Leinaas, J.~Myrheim, Nuovo Cim B \textbf{37}(1), 1 (1977).

\bibitem{wilczek_quantum_1982}
F.~Wilczek, Phys. Rev. Lett. \textbf{49}(14), 957 (1982).

\bibitem{viefers_ideal_1995}
S.~Viefers, F.~Ravndal, T.~Haugset, American Journal of Physics \textbf{63}(4),
  369 (1995).

\bibitem{girardeau_three-body_2010}
M.D. Girardeau, arXiv:1011.2514 [cond-mat]  (2010).

\bibitem{keilmann_statistically_2011}
T.~Keilmann, S.~Lanzmich, I.~McCulloch, M.~Roncaglia, Nature Communications
  \textbf{2}, 361 (2011).

\bibitem{paredes_non-abelian_2012}
B.~Paredes, Phys. Rev. B \textbf{85}(19), 195150 (2012).

\bibitem{lange_strongly_2017}
F.~Lange, S.~Ejima, H.~Fehske, Phys. Rev. A \textbf{95}(6), 063621 (2017).

\bibitem{harshman_one-dimensional_2016}
N.L. Harshman, Few-Body Syst \textbf{57}(1), 11 (2016).

\bibitem{harshman_one-dimensional_2016-1}
N.L. Harshman, Few-Body Syst \textbf{57}(1), 45 (2016).

\bibitem{harshman_integrable_2017}
N.L. Harshman, M.~Olshanii, A.S. Dehkharghani, A.G. Volosniev, S.G. Jackson,
  N.T. Zinner, Phys. Rev. X \textbf{7}(4), 041001 (2017).

\bibitem{harshman_identical_2017}
N.L. Harshman, Few-Body Syst \textbf{58}(2), 41 (2017).

\bibitem{kassel_braid_2008}
C.~Kassel, V.~Turaev, \emph{Braid {Groups}} (Springer, 2008).

\bibitem{harshman_notitle_nodate}
N.~Harshman, A.~Knapp,  In preparation

\bibitem{bjorner_homology_1995}
A.~Bjorner, V.~Welker, Advances in Mathematics \textbf{110}(2), 277 (1995).

\end{thebibliography}
\end{document}